\documentclass[doublecol]{epl2}

\title{Pseudorapidity distributions of produced charged hadrons in $pp$ collisions at RHIC and LHC energies}
\shorttitle{Produced charged hadrons in $pp$ collisions at LHC energies} 

\author{Georg Wolschin}
\shortauthor{G. Wolschin}

\institute{                    
Institut f{\"ur} Theoretische 
Physik
der Universit{\"a}t Heidelberg, 
        Philosophenweg 16,  
        D-69120 Heidelberg, Germany
}

\abstract{The energy dependence of charged-hadron production in proton-proton 
collisions at RHIC and LHC energies is investigated in a nonequilibrium-statistical 
relativistic diffusion model (RDM)
with three sources for particle production. 
Calculated charged-hadron pseudorapidity distributions for
$pp$  at RHIC energies of $\sqrt {s}$ = 0.2 and 0.41 TeV, and  at LHC energies of 0.9, 2.36 and 7 TeV are optimized with respect to the available data. Predictions for 14 TeV are made. The central source arising from gluon-gluon collisions becomes the major origin of particle production at LHC energies. The midrapidity dip is essentially determined by the interplay of the three sources.}

\begin{document}

\maketitle

\newpage
\section{\label{sec:intro}Introduction\protect} 

The investigation of particle production in proton-proton collisions at LHC energies is expected to yield new insights into the underlying partonic processes. Data from the experimental collaborations are now available starting at the injection energy of $\sqrt{s}=0.9$ TeV, via 2.36 TeV, to the current maximum energy of 7 TeV \cite{aamo10,aad10,aadg11,kh10,kha10}. 

A particularly interesting observable is the charged-hadron multiplicity density per unit of pseudorapidity. At midrapidity, it was found to be about 15\% higher than predicted by the available Monte Carlo models that had been calibrated at Tevatron energies \cite{aadg11}.  The distribution functions for non-single-diffractive events have also been measured away from midrapidity, with $|\eta|<2.5$ achieved so far \cite{kha10}. Their shapes are sensitive to the partonic processes that are responsible for charged-hadron production.

In this Letter I propose to analyze the pseudorapidity distribution functions of produced charged hadrons in $pp-$collisions at RHIC and LHC energies as measured by the PHOBOS \cite{alv11} and CMS  \cite{kh10,kha10} collaborations in a schematic
nonequilibrium-statistical model  with three sources.

 Similar ALICE data at LHC energies are also available \cite{aamod10}, as well as older UA5 data \cite{an89} at 0.9 TeV. Corresponding ATLAS results \cite{aad10,aadg11} can not be compared directly with the CMS and ALICE data because particles and events are selected in different regions of phase space. 

The relativistic diffusion model (RDM) has proven to be useful in describing and predicting pseudorapidity distributions of produced charged particles in heavy-ion collsions at SPS, RHIC and LHC energies \cite{gw11}. Related models had also been used in low-energy (non-relativistic) heavy-ion physics \cite{gw82}.

In heavy-ion collisions, the number of produced charged hadrons is much bigger than in $pp$ -- of the order of 20,000 charged hadrons in a central PbPb collision at $\sqrt{s_{NN}}$=2.76 TeV -- and consequently, the application of nonequilibrium-statistical methods such as \cite{wol07,gw11} is clearly justified. 

Special QCD-effects such as the coherence of soft gluons \cite{akr90} that had been predicted by perturbative QCD at low $Q^2$ \cite{erm81,mue81,dok82} to produce visible effects in charged-hadron distributions generated by $e^+e^-$ collisions are less important in the heavy-ion environment since these are averaged out through the random properties of the many-particle system.

Proton-proton collisions at the current maximum LHC energy of 7 TeV produce about 70 charged hadrons integrated over the full rapidity space, including the unmeasured region. Soft-gluon coherence as well as other coherent QCD-effects may still be visible in the data, although less pronounced than in electron-positron collisions. 

The number of produced particles in $pp$ is probably already large enough to test the usefulness of nonequilibrium-statistical concepts, although it may be difficult to observe many-particle effects such as the shift of the fragmentation-peak positions towards midrapidity with decreasing c.m. energy that is clearly seen in the heavy-ion data,  and can be described analytically in the relativistic diffusion model \cite{wol07,gw11}. 

Within the RDM, I investigate in this Letter the energy dependence of the three sources for particle production in proton-proton collisions at RHIC and LHC energies. The energy range considered here covers RHIC energies of $\sqrt {s_{NN}}$ = 0.2 and 0.41 TeV, the presently accessible LHC energies of 0.9, 2.36 and 7 TeV, and the maximum LHC energy of 14 TeV. 

The sources correspond to gluon-gluon induced production of charged hadrons centered at midrapidity, and quark-gluon processes centered at large rapidities, typically $<y_{1,2}> \simeq\mp 2.8$ at 7 TeV. Their relative sizes determine the midrapidity dip in the charged-hadron pseudorapidity distributions when added incoherently. The model is complementary to QCD-based approaches that rely on the corresponding partonic structure functions.

Prominent and detailed models for multiple hadron production are available in the literature. In particular, the dual parton model (DPM) \cite{cap82,cap94} and the equivalent quark-gluon string model \cite{kai82,kma82,kai03} are based on the creation and breaking of quark-gluon strings. 

There the total inclusive hadron production cross section in $pp$ collisions at energies in and below the RHIC energy range arises from contributions of two quark-diquark chains that overlap in rapidity space. These yield the total inclusive cross section when added up incoherently, with a minimum at midrapidity.  At higher energies, also multichain contributions become significant that are likely to contribute in the midrapidity region. There are no transport effects considered in the model.

Although there is presently no direct connection to the DPM, the 3-sources RDM provides an analytical framework to investigate the interplay of central and fragmentation sources, transport effects, and their dependence on incident energy.

The model is considered in Sec. 2, the calculation of pseudorapidity distributions of produced charged hadrons in Sec. 3, and conclusions are drawn in Sec. 4.

\section{Linear Relativistic Diffusion Model}


The Relativistic Diffusion Model (RDM) has been developed to deal with ensembles of many particles and their distribution functions in transverse momentum and rapidity space \cite{wol07}. In particular, it is well-suited to predict and describe charged-hadron rapidity distributions in relativistic heavy-ion collisions from AGS, SPS and RHIC energies, to LHC energies \cite{gw11}. It is tested here for proton-proton collisions at LHC energies, where the number of produced charged hadrons appears to be sufficiently large for nonequilibrium-statistical concepts to apply.

The rapidity distribution of produced particles emerges from an incoherent superposition of the beam-like fragmentation components at larger rapidities arising mostly from valence quark-gluon interactions, and a
component centered at midrapidity that is essentially due to gluon-gluon collisions. All three distributions are broadened in rapidity space as a consequence of diffusion-like processes.


The time evolution of the distribution
functions is governed by a Fokker-Planck
equation (FPE) in rapidity space
\cite{wol07} (and references therein). In the linear model, it is formulated as an
Uhlenbeck-Ornstein \cite{uhl30} process, applied to the
relativistic invariant rapidity for the three components  
$R_{k}(y,t)$ ($k$=1,2,3) of the distribution function
in rapidity space. For a symmetric system such as $pp$ this becomes
\begin{eqnarray}
\lefteqn{
\frac{\partial}{\partial t}R_{k}(y,t)=
-\frac{\partial}
{\partial y}\Bigl[J(y)\cdot R_{k}(y,t)\Bigr]}\nonumber\\&&
\qquad\qquad +\frac{\partial^2}{\partial y^2}
\Bigl[ D_{y}^{k}\cdot R_{k}(y,t)\Bigr].
\label{fpe}
\end{eqnarray}

with the rapidity $y=0.5\cdot \ln((E+p)/(E-p))$. The beam rapidity can also be written as 
$y_{beam}=\mp y_{max}=\mp \ln(\sqrt{s}/m_{p})$.
The rapidity diffusion coefficient $D_{y}$ that contains the
microscopic physics accounts for the broadening of the
rapidity distributions.
The drift function is
 \begin{equation}
J(y)=- y/\tau_{y}
\label{dri}
\end{equation}
with the rapidity relaxation time $\tau_{y}$. It determines the shift of the mean rapidities
towards the central value with increasing time, 

Since the equation is linear, a superposition of the distribution
functions \cite{wol99,wol03} using the initial conditions
$R_{1,2}(y,t=0)=\delta(y\pm y_{max})$
with the absolute value of the beam rapidities 
$y_{max}$, and $R_{3}(y,t=0)=\delta(y)$  
yields the exact solution for a symmetric system.

 In the solution, the mean values 
are obtained analytically from the moments 
equations as
\begin{equation}
<y_{1,2}(t)>=\mp y_{max}\exp{(-t/\tau_{y})}
\label{mean}
\end{equation}
for the sources (1) and (2) with the absolute value of the beam rapidity $y_{max}$.
The gluon-gluon source remains centered at 0 for $pp$ collisions, or other symmetric systems.
Both mean values $<y_{1,2}>$ would attain y=0  
for t$\rightarrow \infty$, whereas for short times they remain between 
beam and equilibrium values. The variances are
\begin{equation}
\sigma_{1,2,eq}^{2}(t)=D_{y}^{1,2,eq}\tau_{y}[1-\exp(-2t/\tau_{y})],
\label{var}
\end{equation}
and the corresponding FWHM-values
are obtained from 
$\Gamma=\sqrt{8\ln2}\cdot \sigma$ since the partial distribution functions are Gaussians in rapidity space
(but not in pseudorapidity space).

The midrapidity source that arises from gluon-gluon interactions with mean value zero comes close to thermal equilibrium with respect to the variable rapidity during the interaction time $\tau_{int}$; the width approaches equilibrium twice as fast as the mean value. I use the notion $R_{gg}(y,t)$ for the
associated partial distribution function in y-space, with 
$N_{ch}^{gg}$ charged particles, cf. table~\ref{tab1}. 

Full equilibrium as determined by the temperature would be reached for $\tau_{int}/\tau_y \gg 1$. The centers of the fragmentation sources would then move to midrapidity according to the solution of the FPE, the incoherent sum of the three sources would reach a thermal distribution in pseudorapidity space. For finite times, however, the fragmentation sources  do not reach $<y_{1,2}>=0$ during the interaction time and hence, remain far from thermal distributions in rapidity space, and do not equilibrate with the central source.

\section{Pseudorapidity distributions}
If particle identification is not available, one has to
convert the results to pseudorapidity, 
$\eta=-$ln[tan($\theta / 2)]$ with the scattering angle $\theta$.
The conversion from $y-$ to $\eta-$
space of the rapidity density
\begin{equation}
\frac{dN}{d\eta}=\frac{dN}{dy}\frac{dy}{d\eta}=\frac{p}{E}\frac{dN}{dy}\simeq
J(\eta,\langle m\rangle/\langle p_{T}\rangle)\frac{dN}{dy} 
\label{deta}
\end{equation}
is performed here through the approximated Jacobian
\begin{eqnarray}
\lefteqn{J(\eta,\langle m\rangle/\langle p_{T}\rangle)=\cosh({\eta})\cdot }
\nonumber\\&&
\qquad\qquad[1+(\langle m\rangle/\langle p_{T}\rangle)^{2}
+\sinh^{2}(\eta)]^{-1/2}.
\label{jac}
\end{eqnarray}

The average mass $<m>$ of produced charged hadrons in the
central region is approximated by the pion mass $m_{\pi}$ since pions represent by far the largest fraction of produced charged hadrons, in particular in the midrapidity source where the transformation has the
biggest effect. The mean mass is larger in the fragmentation region, with $<m> \simeq m_p/n_{ch}^{1,2}+m_{\pi}\cdot(n_{ch}^{1,2}-1)/n_{ch}^{1,2} \simeq 0.27$ GeV at $\sqrt{s}=7$ TeV where $n_{ch}^{1,2} = dN/d\eta (<\eta_{1,2}>) \simeq 6$.

Due to the Jacobian, the partial distribution functions differ from Gaussians, but as a consequence of the relatively high mean transverse momenta at RHIC and, in particular, LHC energies $<p_{T}>=0.39-0.61$ GeV (see table~\ref{tab1} and \cite{kha10}) the Jacobian has only a very small effect on the central source at sufficiently high values of $\sqrt{s}$, and almost no effect on the fragmentation sources.

\begin{figure}
\includegraphics[width=8.8cm]{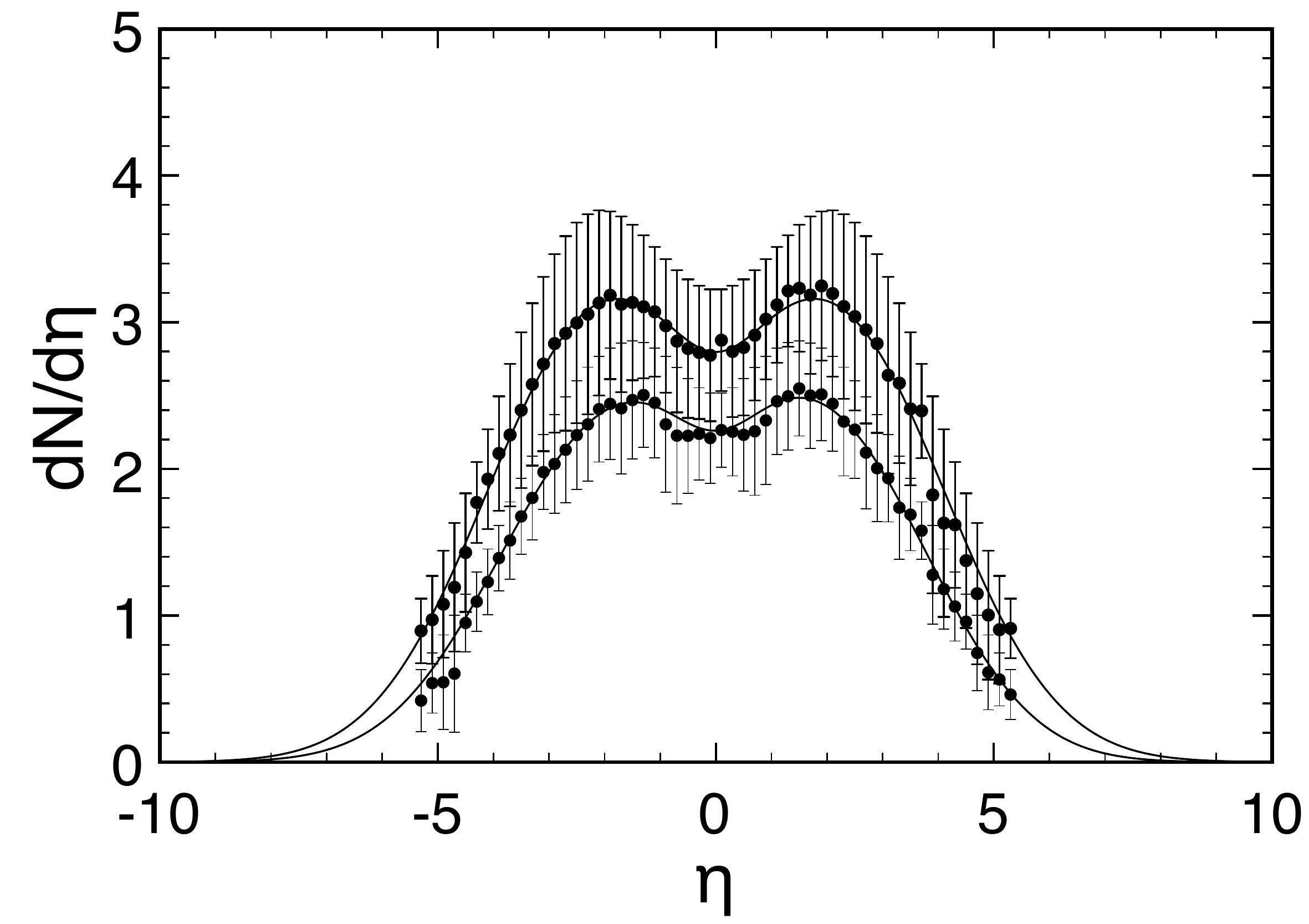}
\caption{Pseudorapidity distributions of produced charged hadrons in inelastic $pp$ collisions at RHIC energies
 of $\sqrt{s}=$ 0.2 and 0.41 TeV as calculated with the three sources and fitted to PHOBOS data \cite{alv11}.} 
\label{fig1}
\end{figure}

\begin{figure}
\includegraphics[width=8.8cm]{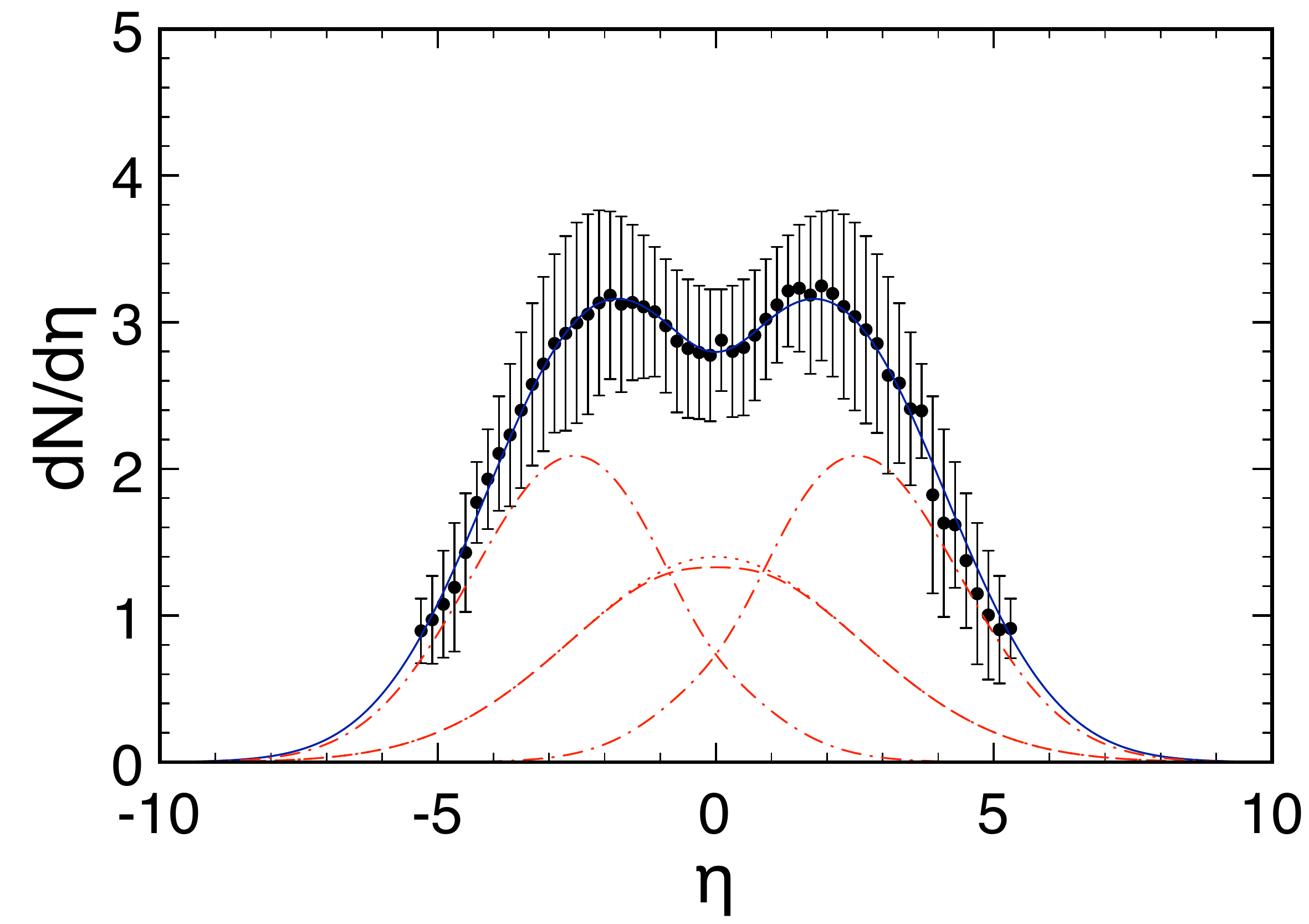}
\caption{Pseudorapidity distributions of produced charged hadrons in inelastic $pp$ collisions at RHIC energies
 of $\sqrt{s}= $0.41 TeV as calculated with the three sources and fitted to PHOBOS data \cite{alv11}. 
The underlying partial distribution functions including the Jacobian are shown as dash-dotted curves for the fragmentation sources arising from quark-quark and quark-gluon interactions, and as dashed curve for the central source arising from gluon-gluon interactions. The dotted curve does not include the Jacobian.} 
\label{fig2}
\end{figure}

\begin{figure}
\includegraphics[width=8.8cm]{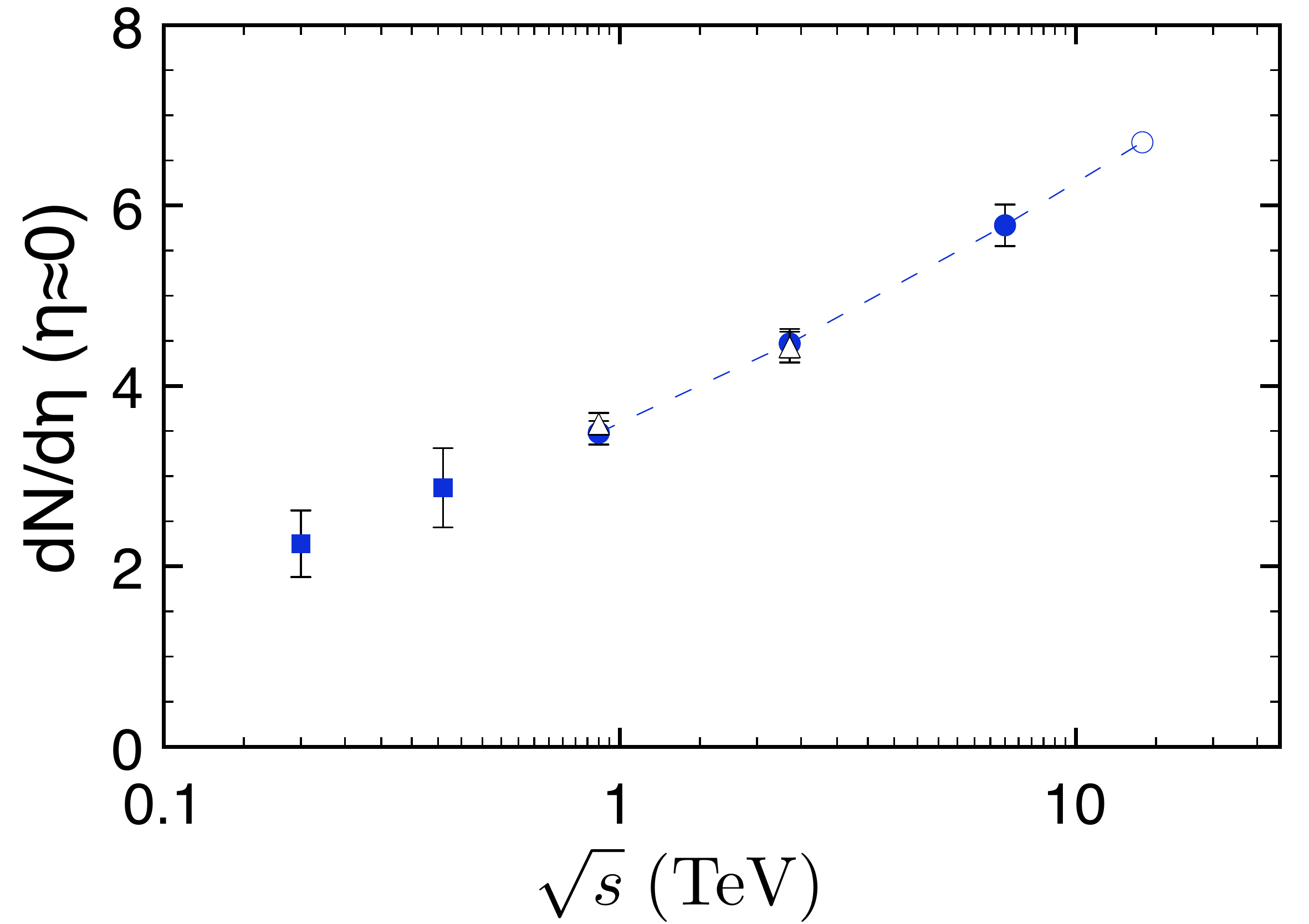}
\caption{Charged-particle pseudorapidity densities in the central pseudorapidity region $|\eta|<0.5$ for non-single-diffractive (NSD) proton-proton collisions as function of the centre-of-mass energy $\sqrt{s}$. The value at 14 TeV (circle) is extrapolated from CMS data 
(dots, \cite{kha10}). ALICE NSD data at 0.9 and 2.36 TeV are shown for comparison (triangles, \cite{aamod10}).
Squares at RHIC energies of 0.2 and 0.41 TeV are inelastic PHOBOS $pp-$data for $|\eta|<1$ \cite{alv11}.}
\label{fig3}
\end{figure}

\begin{figure}
\includegraphics[width=8.7cm]{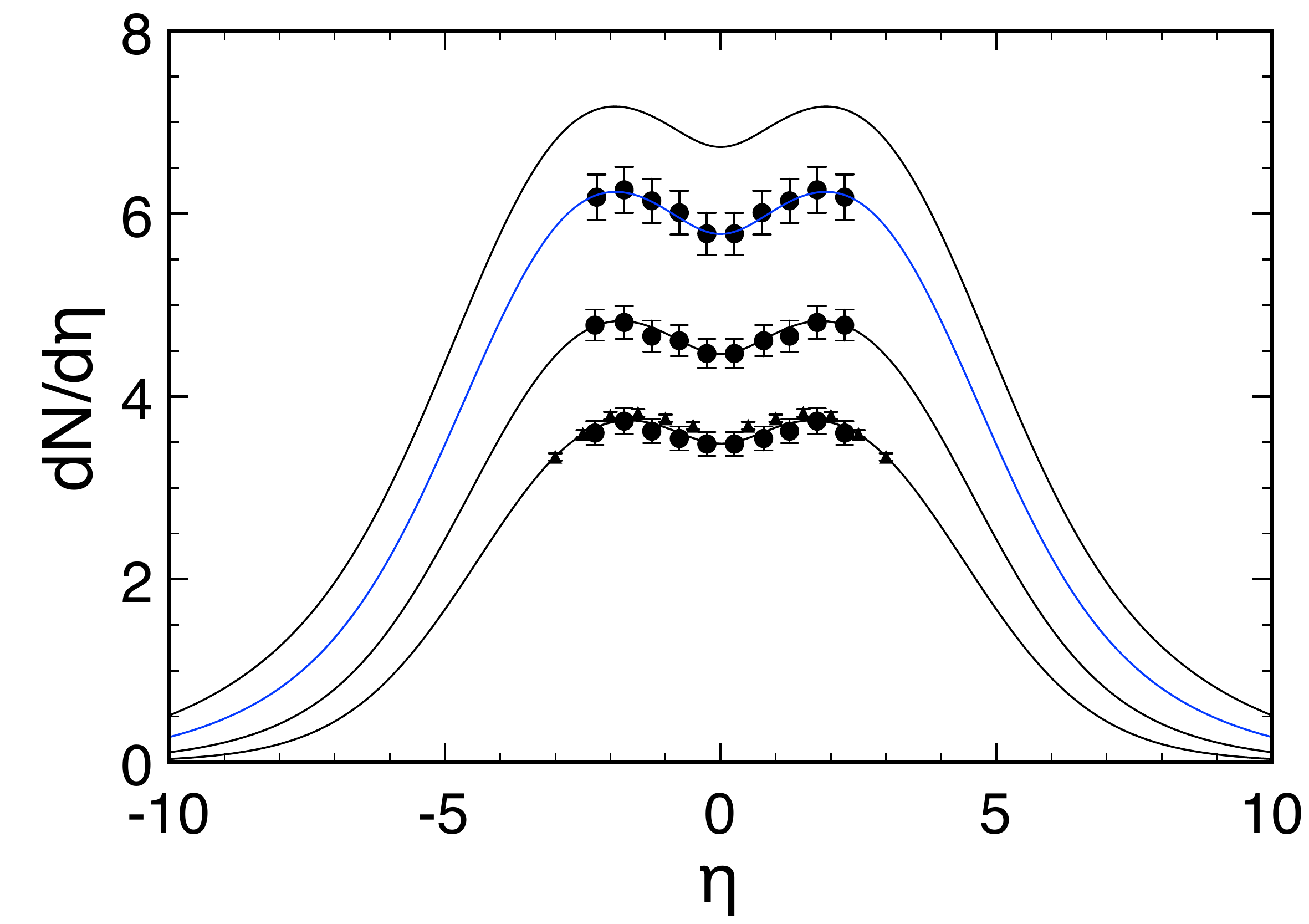}
\caption{Pseudorapidity distributions of produced charged hadrons in $pp$ collisions (NSD) at LHC c.m. energies of 0.9, 2.36, 7 and 14 TeV (bottom to top) as calculated in the three-sources approach and fitted to CMS NSD data \cite{kh10,kha10}.  At 0.9 TeV UA5 NSD data are also shown \cite{an89}, triangles. See fig.~\ref{fig5} for the underlying partial distribution functions at 7 TeV.}
\label{fig4}
\end{figure}

\begin{figure}
\includegraphics[width=8.7cm]{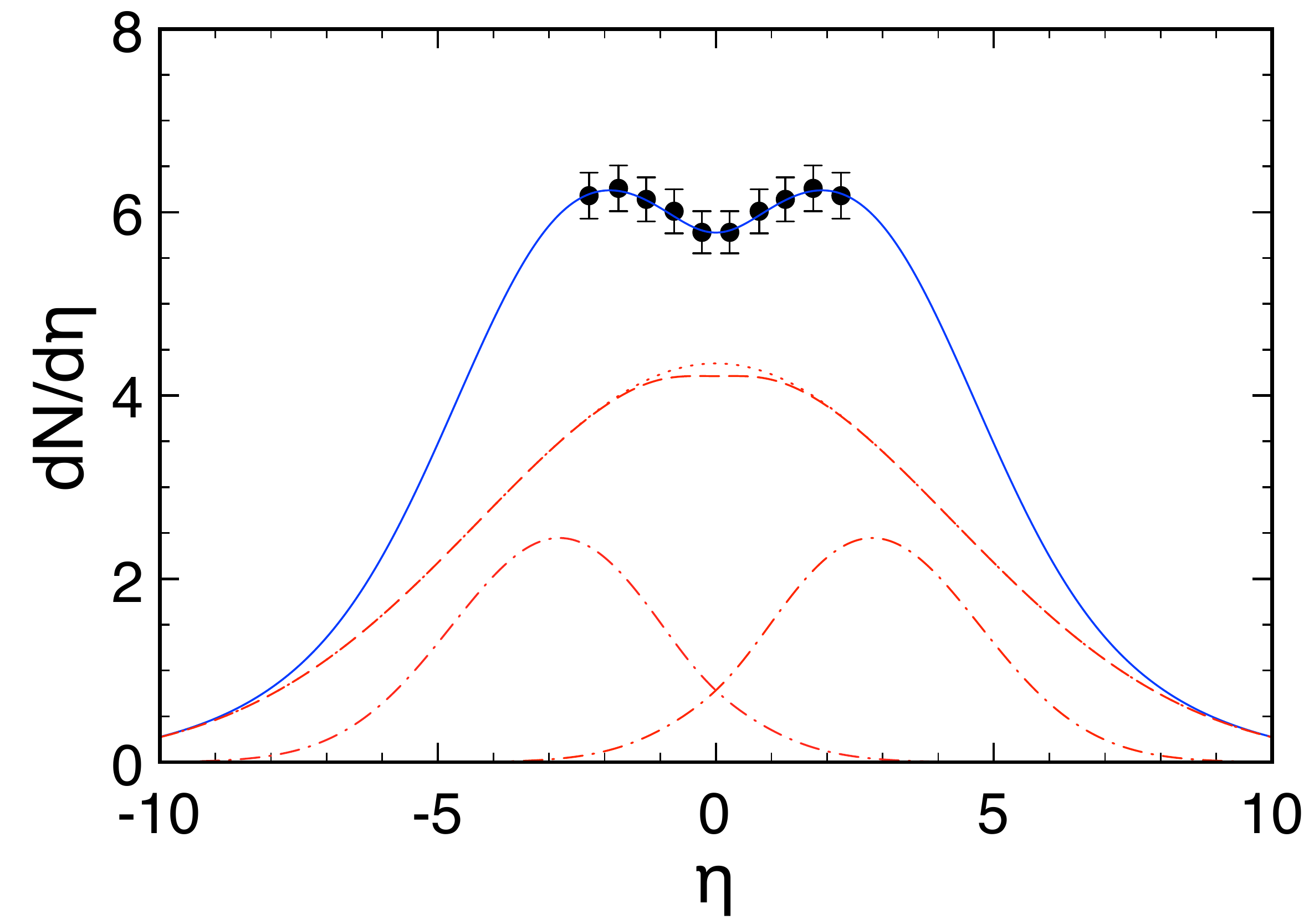}
\caption{Pseudorapidity distributions of produced charged hadrons in $pp-$collisions at LHC c.m. energy of  7 TeV.
The underlying partial distribution functions including the Jacobian are shown as dash-dotted curves for the fragmentation sources arising from quark-quark and quark-gluon interactions, and as dashed curve for the central source arising from gluon-gluon interactions. The dotted curve does not include the Jacobian.}
\label{fig5}
\end{figure}

\begin{table*}
\begin{center}
\caption{\label{tab1}Three-sources parameters for $pp$ collisions at RHIC energies (upper two lines) and  at LHC energies (lower four lines).
At RHIC energies the fragmentation sources from quark-gluon interactions with particle content $N_{ch}^{1,2}$ dominate. At LHC energies the source from gluon-gluon collisions with particle content $N_{ch}^{gg}$ is the major origin of particle production at midrapidity.
Midrapidity values (last column) are from PHOBOS (inelastic) \cite{alv11} for $|\eta| < 1$ at RHIC energies,
and from CMS (NSD) \cite{kh10,kha10}  for $|\eta| < 0.5$ at LHC energies. The 14 TeV value is calculated with the extrapolated parameters. See \cite{kha10} for approximate average $<p_T>-$values.}
\vspace{.6cm}
\begin{tabular}{llllllllcr}
\hline\\
$\sqrt{s} $&$y_{beam}$& $<p_T>$&$\tau_{int}/\tau_y$&$<y_{1,2}>$&$\Gamma_{1,2}$&$\Gamma_{gg}$&$N_{ch}^{1,2}$&$N_{ch}^{gg}$&$\frac{dN}{d\eta}|_{\eta \simeq 0}$\\
   (TeV)&&(GeV/c)\\
\hline\\
     
  0.20&$\mp 5.362$&0.39&0.85&$\mp 2.30$&4.4 &4&9&4&$inel \hspace{.2cm} 2.25^{+0.37}_{-0.30}$\cite{alv11}\\
 0.41&$\mp 6.080$&0.42&0.89&$\mp 2.50$&4.5&10&10&9&$inel\hspace{.2cm} 2.87^{+0.44}_{-0.43}$\cite{alv11}\\
&&&&&&&&\\
 0.90&$\mp 6.866$&0.46&0.93&$\mp 2.70$&4.6&8&8&21&3.48$\pm 0.02\pm 0.13$ \cite{kh10}\\
 2.36&$\mp 7.830$&0.50&1.05& $\mp 2.75$&4.6&9&10&31&4.47$\pm 0.04\pm 0.16$ \cite{kh10}\\
  7.00&$\mp 8.918$&0.55&1.16& $\mp 2.80$&4.6&10&12&46&5.78$\pm 0.01\pm 0.23$\cite{kha10}\\
  14.00&$\mp 9.611$&0.61&1.22& $\mp 2.85$&4.8&11&14&59&6.73$\pm 0.30$\\\\

\hline
\end{tabular}
\end{center}
\end{table*}

For heavy-ion systems, the dependencies of the diffusion-model parameters on incident energy, mass and centrality at RHIC  and LHC energies have been investigated in \cite{wob06,kw07,kwo07, gw11}.
This Letter presents the first investigation within the RDM for $pp$ at high relativistic energies.
The corresponding parameters are shown in table~\ref{tab1} as functions of the c.m. energy $\sqrt{s}$.

The time parameter
$\tau_{int}/\tau_{y}$ is displayed as function of center-of-mass energy in
table~\ref{tab1}. It is seen to increase with $\sqrt{s}$.
An increasing time parameter implies that the local maxima of the distribution function move further away from the beam rapidity (that increases with $\ln(\sqrt{s}/m)$) with increasing energy. In accordance with the expectation, it indicates that the rapidity equilibration time $\tau_y$ decreases with rising energy, whereas the interaction time $\tau_{int}$ depends only weakly on energy in the $pp$ system. 

From the available data, it appears that the local maxima occur at rather similar positions in pseudorapidity space, $\eta \simeq 2$. This is characteristically different from heavy-ion collisions, where the maxima move outwards with increasing energy, as observed in AuAu data at RHIC \cite{alv11},
and described in the RDM \cite{gw11}. This qualitative difference is most likely due to the larger spatial extent of the heavy-ion system.

The partial widths (FWHM) as functions of energy are found to increase linearly with $\log{\sqrt{s}}$, table~\ref{tab1}.  
Here the widths are effective values: beyond the statistical widths
that can be calculated from a dissipation-fluctuation theorem \cite{wols99} within the RDM,
they include the effect of collective expansion of the produced particles.
The values at RHIC energies are resulting from a minimization with respect to the data that corresponds to the
time evolution up to $\tau_{int}$. The integration is 
stopped at the optimum values of 
$\tau_{int}/\tau_{y}$, $\Gamma_{1,2,gg}$, and $N_{ch}^{gg}$ and hence,
the explicit value of $\tau_{int}$ is not needed.

The normalization is given by the total number of produced charged hadrons that is taken from experiment if available, or extrapolated in case of predictions at higher energies. Hence, the model contains four parameters. It 
provides an analytical framework to calculate the distribution functions, and to draw physical conclusions.


The charged-particle distributions in rapidity space are obtained
as incoherent 
superpositions of nonequilibrium and central (``equilibrium") solutions of
 (\ref{fpe}) 
\begin{eqnarray}
    \lefteqn{
\frac{dN_{ch}(y,t=\tau_{int})}{dy}=N_{ch}^{1}R_{1}(y,\tau_{int})}\nonumber\\&&
\qquad\qquad +N_{ch}^{2}R_{2}(y,\tau_{int})
+N_{ch}^{gg}R_{3}(y,\tau_{int}).
\label{normloc1}
\end{eqnarray}


\section{Results and discussion}

The results for pseudorapidity distributions of produced charged 
hadrons in inelastic $pp$ collisions at two RHIC energies of 0.2 and 0.41 TeV are shown in fig.~\ref{fig1}
in comparison with PHOBOS data \cite{alv11}. The three-sources model yields excellent agreement with the data. Here the overall normalization is taken from the data, and the fit parameters are the time parameter (that determines the mean values $<y_{1,2}>$), the widths $\Gamma_{1,2}, \Gamma_{gg}$, and the number of produced particles in the central source $N_{ch}^{gg}$.

At RHIC energies, the multiplicity density at midrapidity has still a substantial contribution from the overlapping fragmentation sources. At 0.2 TeV, the contribution from the gluon-gluon source at $\eta=0$ is about 20 \%, at 0.41 TeV the midrapidity source is already much larger (48\%), but the fragmentation sources still contribute 26\% each, as shown in fig.~\ref{fig2}.

It should be mentioned that there exist detailed microscopic calculations of fragmentation sources from 
$gq \rightarrow q$ and $qg \rightarrow q$ diagrams by Szczurek et al. \cite{sz04,csz05} for pion production in proton-proton and heavy-ion collisions at SPS and RHIC energies. These processes are also responsible for the observed differences \cite{bea01} in the production of positively and negatively charged hadrons, in particular, pions. An extension of these calculations to LHC energies is very desirable.

To determine the RDM parameters in $pp$ collisions at LHC energies, I have performed
fits of the time parameters to the maxima of the double-humped charged-hadron distributions, extrapolations of the partial widths $\Gamma_{1,2,gg}$ with $\log{\sqrt{s}}$, and corresponding extrapolations of the number of produced particles in fragmentation and central sources as functions of $\log{\sqrt{s}}$, see table~\ref{tab1}. The number of particles in the central sources is at sufficiently high energy essentially determined by the measured pseudorapidity density near midrapidity that is plotted in fig.~\ref{fig3} as function of energy.

The results at LHC energies are shown in fig.~\ref{fig4}.  The model results are compared with CMS data at 0.9, 2.36 and 7 TeV \cite{kh10,kha10}, and UA5 data at 0.9 TeV \cite{an89}.
The calculation at 14 TeV is performed based on an extrapolation of the multiplicity density at midrapidity with $\log{\sqrt{s}}$ that yields $dN/d\eta \simeq 6.73 \pm 0.30$ at midrapidity. 

At LHC energies, the overall scenario changes in favor of particle production from the midrapidity source. 
The bulk of the midrapidity density is generated in the central source (73\%) at 7 TeV,
there is only a small overlap of the fragmentation sources at midrapidity as shown in fig.~\ref{fig4}.

In a comparison with calculations at LHC energies that do not include the Jacobian transformation as displayed by the dotted curve in fig.~\ref{fig5}, it is evident that the midrapidity dip structure is essentially determined within the RDM by the interplay of the three sources for particle production, and only marginally influenced by the transformation from $y-$ to $\eta-$space at these high energies. The central distribution including the Jacobian has no dip at LHC energies, but only a slight reduction in absolute magnitude at midrapidity, as shown by the dashed curve in fig.~\ref{fig5}. 

There is, however, also the possibility that coherent QCD-effects contribute to the dip structure. Such effects go beyond the present calculation. They are also not considered in numerical event generators (see \cite{kw06} as an example), which provide rather accurate representations of RHIC data, in particular for transverse momentum distributions. As compared to the analytical RDM these require, however, a substantial numerical effort. 

Another purely empirical formulation of pseudorapidity distributions in multiple particle production at $\sqrt{s}=$ 22.4 to 1800 GeV based on several emitting centers along the rapidity axis had been given in \cite{ost10}. It yields an analytical expression for the distribution function, and four parameters are fitted to the data. As compared to the straightforward physical interpretation of the three sources in the RDM it seems, however, difficult to assign a physical meaning to the sources.

The determination of the parameters within the RDM clearly goes beyond triple-gaussian fits that are modified by the Jacobian, because the comparison with the data is based on, and constrained by, the underlying nonequilibrium-statistical description. Hence the dependence of the resulting parameters on incident energy as shown in table 1 is not arbitrary, but yields a consistent physical result.

In particular, the time parameter increases with center of mass energy since the rapidity relaxation time decreases. The width and particle content of the fragmentation sources do not change much with rising energy because the number of contributing valence quarks stays constant, whereas the width and, in particular, the particle content of the central source that arises from gluon-gluon collisions increases substantially due to the large increase of gluons in the system at high energy and small values of Bjorken-$x$.

As compared to the application of the RDM to heavy-ion collisions, it appears that transport phenomena are not fully developed in $pp$ due to the small transverse size of the system.
In the energy range from 0.2 to 14 TeV considered here, the peak positions stay almost constant at $\eta \simeq 2$ in pseudorapidity space. The shift with energy that is present in $AA$ systems \cite{kw07}, and interpreted there as a multiparticle effect, does not seem to occur in $pp$. Hence, the full development of transport phenomena in higly relativistic collisions requires
a sufficiently large system in transverse size.

\section{Conclusion}
Based on the description of charged-hadron pseudorapidity distributions
in $pp$ collisions at RHIC and LHC energies in a nonequilibrium-statistical model,
I have presented calculations of pseudorapidity distributions of produced charged hadrons for $pp$ collisions at RHIC energies of 0.2 and 0.41 TeV, and at LHC energies of 0.9, 2.36, 7 and 14 TeV.
These rely on the extrapolation of the transport parameters in the relativistic diffusion model (RDM) with increasing center-of-mass energy, and fits to the available data.

In a three-sources model, the midrapidity source that is associated with gluon-gluon collisions accounts for about 73\% of the charged-particle multiplicity density measured by CMS at midrapidity in $pp$ collisions at 7 TeV. The fragmentation sources that correspond to particles that are mainly generated from valence quark -- gluon interactions are centered at relatively large values of pseudorapidity $(<\eta_{1,2}>\simeq<y_{1,2}> \simeq \mp 2.8)$ and hence, these contribute only marginally to the midrapidity yield.

Since the Jacobian transformation from rapidity to pseudorapidity space is close to 1 at LHC energies due to the large mean transverse momenta, the size of the midrapidity-dip in the pseudorapidity distribution function is essentially determined by the relative particle content in the three sources, not by the Jacobian. Small corrections of the extrapolated values for the number of produced particles in the fragmentation sources may be required should measured distributions beyond pseudorapidity $\eta =2.5$ become available from CMS, ATLAS and ALICE at LHC energies.\\

\bf{Acknowledgments}

\rm
This work has been supported
by 
the ExtreMe Matter Institute EMMI.
\bibliographystyle{eplbib}
\bibliography{gw_epl}


\end{document}